# Polarons in condensed matter

**David Emin**, Department of Physics and Astronomy, University of New Mexico, Albuquerque, NM 87131, USA

emin@unm.edu

## ABSTRACT

Polarons are composite quasiparticles comprising electronic charge carriers taken together with the alterations they induce in surrounding condensed matter. Strong-coupling polarons form when electronic charge carriers become 'self-trapped:' bound within potential wells stabilized by carriers' presence. Distinctively, exciting these bound carriers generates broad absorption bands. Strong-coupling polarons are slow and massive since moving them requires atomic motion. Their transport differs qualitatively from that of conventional electronic charge carriers. Large (strong-coupling) polarons move coherently with mobilities that fall with rising temperature. These massive quasiparticles' very weak scatterings by phonons produce much lower room-temperature mobilities than those permitted of conventional electronic charge carriers. Moreover, the long scattering times associated with large polarons' weak scattering relegates their principal ac (Drude) transport to below phonon frequencies. Small (strong-coupling) polarons move incoherently with even lower thermally assisted mobilities. Strikingly, a magnetic field often deflects small polarons in the opposite sense than it does conventional charge carriers, thereby producing anomalously signed Hall Effects. In exceptional circumstances charge carriers self-trap in pairs thereby forming large and small bipolarons. Some transport features distinguish them from polarons. Interference between carrier-induced atomic displacement patterns produce attractive interactions between like-charged polarons and short-range repulsive interactions between oppositely charged polarons.

Key points:

- An electronic carrier is self-trapped when it is bound in a potential well produced by atoms assuming displaced equilibrium positions stabilized by the carrier's presence.

- Displaced distant ions generate the potential well which binds a large strong-coupling polaron's multi-site self-trapped carrier. Displacements of nearby atoms create the potential well which drives the collapse of a small strong-coupling polaron's self-trapped carrier to a single site. Disorder fosters small-polaron formation.

- Unusual transport and optical properties of large and small polarons distinguish them from each other and from conventional electronic carriers.

- Polarons tend to form in unusually complex materials. Polarons' distinctive properties enable unexpected applications.

## Introduction

Adding electronic charge carriers to condensed matter generally affects its atoms' motion. A polaron is defined as the composite quasiparticle composed of an electronic carrier plus the altered atomic motion its presence induces.

There are two regimes in which polaron effects have been studied. In the weak-coupling regime, the relaxation of the electronic charge carrier in response to atomic movements only alters their frequencies. An electronic carrier then lowers the ground-state energy of the interacting system by a modest amount that vanishes as atomic vibrations are suppressed. Much larger energy shifts occur in the strong-coupling regime. Then the electronic carrier is 'self-trapped:' bound within the potential well produced by carrier-induced shifts of atoms' equilibrium positions. A dynamically stable strong-coupling polaron forms when the binding energy of the self-trapped electron greatly exceeds the characteristic phonon energy. Carrier-induced reductions of the frequencies with which atoms vibrate about their displaced equilibrium positions then make subsidiary contributions to a strong-coupling polaron's energy.

Unlike weak-coupling polarons, strong-coupling polarons have electronic and optical properties which distinguish them from conventional electronic charge carriers. Henceforth, this article will only consider strong-coupling polarons.

Self-trapping, the distinctive feature of strong-coupling polarons, is a nonlinear phenomenon. The potential well within which a self-trapped electronic carrier is bound results from atoms' assuming displaced equilibrium positions. However, these displacements of atoms' equilibrium positions are driven by the presence of the bound electronic carrier. That is, the potential well which binds a strong-coupling polaron's electronic carrier depends on its wavefunction. Solution of the carrier's nonlinear wave-equation generally permits two distinct types of strong-coupling polaron. A large-polaron's self-trapped carrier extends over multiple structural units. By contrast, a small-polaron's electronic carrier collapses to its minimal volume, e.g. a single structural unit.

A polaron's formation is governed by the forces between its electronic charge carrier and surrounding displaceable atoms. These interactions are often termed electron-lattice or electron-phonon interactions. A large polaron's formation usually requires long-range Coulomb-based interactions (e.g. the Fröhlich interaction) between its electronic carrier and condensed-matter's distant displaceable ions. By contrast, a small polaron's formation is usually generated by the energy of its electronic carrier depending on separations between the atoms it contacts. Separate large-polaron and small-polaron solutions can occur in the general case in which the electron-phonon interaction is the sum of long-range and short-range components.

The ratio of the static to optical dielectric constant, $\varepsilon_0/\varepsilon_\infty$, measures the prevalence of displaceable ions within a semiconductor. In particular, a ratio comparable to 1 indicates a covalent material with its purely short-range electron-phonon interaction. By contrast, ratios greater than or comparable to 2 (the value typifying alkali halides) indicate sufficiently displaceable ions to support the long-range electron-phonon interaction required for large-polaron formation. Ratios that greatly exceed 2 open the possibility of exceptional polaron effects, such as self-trapped carriers pairing to thereby form bipolarons.

A large polaron's self-trapped electronic carrier extending over multiple sites insures that its motion between adjacent sites is also coherent. Unlike the coherent transport of a conventional electronic charge carrier, a large polaron's motion is contingent on the movement of atoms. Therefore, a large polaron's motion is orders-of-magnitude slower and its effective mass is orders-of-magnitude greater than those of conventional electronic charge carriers. As a result, coherent large-polaron transport occurs with much lower mobilities than can be ascribed to conventional electronic charge carriers. In particular, the minimum mobility possible for a carrier's coherent transport occurs when its mean-free-path falls to its diameter, its deBroglie wavelength. This value is $qh/m_c kT$, where $q$ and $m_c$ denote the magnitudes of a carrier's charge and effective mass, $T$ represents the absolute temperature while $h$ and $k$ respectively signify the Planck and Boltzmann constants. This minimum mobility is 300 cm$^2$/V-s at room temperature for a carrier with a free electron's charge and mass. Thus, smaller room-temperature mobilities than this are inconsistent with the transport of conventional electronic charge carriers. However, such small mobilities are compatible with those of large polarons. For example, the mobility of a large polaron scattered by acoustic phonons is $qc_sR_{lp}/kT$ above a fraction of the Debye temperature, where $c_s$ denotes the sound velocity and $R_{lp}$ represents the large-polaron radius.

The distinctively extreme localization of a small-polaron's self-trapped carrier generates incoherent charge transport. Small polarons then move by successions of jumps between sites. The resulting small-polaron mobility is extremely small, $<< 1$ cm$^2$/V-s, and rises with increasing temperature. This thermally assisted hopping is Arrhenius above a significant fraction of the characteristic phonon temperature. A small-polaron's mobility manifests a progressively weaker temperature dependence as the temperature is lowered.

Large and small polarons also manifest optical properties which distinguish them from one another and from conventional electronic charge carriers. Unlike conventional charge carriers, large and small polarons both generate very broad asymmetric absorption bands that peak at energies well above phonon energies. A large-polaron's broad absorption band is generated by photo-ionizing its self-trapped electronic charge carrier from the potential well within which it is bound. With increasing photon energy, a large-polaron absorption band rises toward its peak and diminishes more slowly thereafter. A small-polaron's absorption of a photon induces its self-trapped carrier to hop. Each absorbed photon produces a small-polaron jump. With rising temperature, increased atomic vibrations increase the ranges of initial and final-state electronic energies involved in these photo-induced transitions. The width of a small-polaron absorption band is thereby thermally broadened. With increasing photon energy, a typical small-polaron absorption band rises toward its peak and then diminishes relatively rapidly. Thus, the thermally broadened asymmetry of a small-polaron absorption band distinguishes it from a large-polaron absorption band.

Large-polarons' coherent motion also produces an absorption analogous to the Drude absorption of conventional electronic charge carriers. Distinctively, a large polaron's huge mass causes it to be only very weakly scattered by ambient phonons that reflect off its softened atomic vibrations. As such, a large-polaron's motion is characterized by an extremely long relaxation time (proportional to its mass) that limits its Drude-like contribution to below the characteristic phonon frequency. Thus, there is a pseudo-gap in a large-polaron's absorption spectrum between its very low-frequency Drude-like absorption and the onset of its broad high-frequency photo-ionization band. This pseudo gap opens with decreasing temperature as phonon scattering is progressively suppressed.

General features governing the formation and properties of strong-coupling polarons were succinctly described in this introductory section. However, other dramatic manifestations of the formation and transport of strong-coupling polarons rely on specific features of the condensed matter in which they form. For example, the controlled imposition of disorder can induce charge carriers' abrupt transformation between their being conventional charge carriers or large polarons and being small polarons. Unlike conventional carriers, small-polarons often form in materials whose local site arrangements generate Hall Effect sign anomalies. Small polarons are then deflected by a magnetic field in the opposite sense to that of free particles bearing the same charge. The self-trapping of holes on oxides' oxygen-related states transfers electrons to surrounding ions. As such, formation of such polarons can alter magnetic moments of surrounding magnetic ions. Special features emerge in materials containing especially displaceable ions as manifested by exceptionally large ratios of their static to high-frequency dielectric constants. These situations favor attractive interactions among large polarons inducing their pairing into bipolarons. Furthermore, a gas of large bipolarons can even condense into a large-bipolaron liquid. Short-range repulsive interactions between polarons of opposite charge suppress their recombination thereby increasing the lifetimes of optically induced polarons. These and other topics will be addressed in the following sections.

## Long-range electron-phonon interactions

Polaron effects address the responses of electronic charge carriers to the comparatively slow movements of condensed-matter's relatively massive atoms. These responses are driven by electron-phonon interactions. Electron-phonon interactions arise from dependences of an electronic charge carrier's potential energy on the positions of the nuclei of the atoms within which it is immersed.

An electronic carrier's Coulomb interactions with ions that surround it generate a long-range component of its electron-phonon interactions. A simple model for the long-range component considers only ions that are outside of the electronic carrier's radius. Treating the condensed medium as a continuum, the polarizability associated with displacing nuclei is represented as the difference between the continuum's net polarizability and that due to its bound electrons. As such, the effective coupling constant considers a Coulomb interaction modified by the factor $1/\varepsilon_\infty - 1/\varepsilon_0$, the difference between the reciprocals of the high-frequency dielectric constant $\varepsilon_\infty$ and the static dielectric constant $\varepsilon_0$. This model's coupling strength is the Fröhlich constant, $\alpha \equiv e^2(1/\varepsilon_\infty - 1/\varepsilon_0)(m_e/2\hbar^3\omega)^{1/2}$, where $e$ and $m_e$ respectively denote the electronic carrier's charge and effective mass with $\omega$ signifying an optic-mode atomic vibration frequency. This continuum theory addresses a very large-radius self-trapped state with $\alpha^2 \hbar\omega$ essentially being its binding energy. This model has been applied to self-trapped electrons in alkali halides.

Electrons in polar liquids such as water and ammonia are self-trapped. In these instances, the liquids' polar molecules are oriented to produce the potential well which binds the self-trapped electrons. Thus, for example, water molecules' electropositive hydrogen atoms are closer to the self-trapped electron than are their electronegative oxygen atoms.

## Short-range electron-phonon interactions

The electron-phonon interaction in covalent semiconductors, termed the deformational-potential, is short-range. In particular, a charge carrier's energy depends on the separations between the atoms linked by the bonding or anti-bonding state that it occupies. The dependence of an electronic carrier's energy on the inter-atomic separation is typically quite large, 2-3 eV/Å. For comparison, the Coulomb interaction between two elementary charges separated by 3 Å is less than 5 eV even when the reduction associated with the dielectric constant (12 and 16 in Si and Ge, respectively) is ignored.

The short-range electron-phonon interaction's effect on an electronic state is inversely proportional to the number of atoms or bonds it contacts. Thus, for example, the electron-phonon interaction in a covalent semiconductor has a very much smaller effect on a large-radius donor state (e.g. a P donor state in Si) than it does on a well-localized donor state. Similarly, the effect of the short-range electron-phonon interaction on an electronic carrier in a molecule tends to decrease as its size increases.

A short-range electron-phonon interaction can predominate even in ionic solids such as alkali halides. For example, a hole in the alkali halide KCl is shared between two nearest-neighbor Cl anions to produce the molecular anion $(Cl_2)^-$: hole $+ 2Cl^- \rightarrow (Cl_2)^-$. In other words, this self-trapped hole is severely localized in the chemical bond which forms between two displaced Cl anions.

Charge transfer between oxygen anions and surrounding cations generates a distinctive short-range electron-phonon interaction. Despite an isolated oxygen atom only having an electron affinity for one electron, oxygen anions are widely represented as having an oxidation state of −2. That is, the cations that typically surround an oxygen anion in a solid facilitates its accommodating an extra electron. This electron will leave an oxygen anion as surrounding cations are displaced away from it.

Cations in ionic solids sometimes possess magnetic moments. Without charge carriers, super-exchange between these magnetic cations usually drives these solids to form anti-ferromagnetic insulators. Dopants and defects can introduce charge carriers. Intra-atomic exchange interactions between charge-carriers' spins and the cations' magnetic moments foster their ferromagnetic alignment. A carrier-induced ferromagnetic cluster within an antiferromagnetic insulator is termed a magnetic polaron. Although the electronic carrier bound within a prototypical magnetic polaron (e.g. an electron in EuTe) may be only mildly localized, electron-phonon interactions will increase its localization.

## Self-trapping: Formation of a strong-coupling polaron

A strong-coupling polaron is defined by having its electronic charge carrier self-trapped. A self-trapped electronic carrier is bound within the potential well generated by surrounding atoms assuming equilibrium positions consistent with the electronic carrier occupying its bound state. Moving such a polaron requires displacing atoms from their carrier-induced equilibrium positions. The validity of this approach is generally taken to require that the binding energy of the self-trapped electronic carrier $-E_{st}$ exceed the phonon energy characterizing associated atoms' vibrations $\hbar\omega$.

The carrier-induced displacements of atoms' equilibrium positions are found by minimizing the Born-Oppenheimer adiabatic potential. This potential energy is the sum of 1) the potential energy associated with displacing atoms from their carrier-free equilibrium positions plus 2) the electronic energy as a function of atoms' positions. The adiabatic potential with atoms at these displaced equilibrium positions establishes the potential well within which an electronic carrier is self-trapped.

The chemistry underlying these interactions can be complex. Nonetheless, here, as is typical, electronic-carrier's energies and inter-atomic strain energies are simply modelled to respectively have linear and quadratic dependences on atoms' displacements from their carrier-free equilibrium values.

A general approach to self-trapping proceeds by treating the solid as a deformable continuum. The ground-state wave function for the self-trapped electronic carrier $\varphi(\boldsymbol{r})$ with energy $E_{st}$ is then the solution of the non-linear wave equation:

$$\left[\frac{-\hbar^2}{2m_e}\nabla_r^2 - \frac{1}{S}\int d\boldsymbol{r}'|\varphi(\boldsymbol{r}')|^2 \int d\boldsymbol{u}\, F(\boldsymbol{r}-\boldsymbol{u})F(\boldsymbol{r}'-\boldsymbol{u})\right]\varphi(\boldsymbol{r}) = E_{st}\varphi(\boldsymbol{r}). \quad (1)$$

Here $m_e$ denotes the electronic carrier's effective mass and $S$ represents the continuum's stiffness constant per unit volume. The electron-phonon force per unit volume $F(\boldsymbol{r}-\boldsymbol{u})$ describes how the electronic potential energy at location $\boldsymbol{r}$ depends on a fiducial deformation of the continuum at location $\boldsymbol{u}$.

This non-linear wave equation can be utilized to express the self-trapping energy as a functional of its wavefunction. This functional is then recognized to depend simply on the dimensionless radius $R$ of the self-trapped electron's normalized wavefunction. For a charge carrier which can move among sites with dimensionality $d$ embedded within a three-dimensional ionic medium:

$$E_{st}(R) = \frac{T_e}{R^2} - \frac{2V_S}{R^d} - \frac{2V_L}{R}. \quad (2)$$

Here $T_e$, $V_S$ and $V_L$ are constants related to the self-trapped electronic carrier's kinetic energy and contributions to its potential energy from the short-range and long-range components of its electron-phonon interaction. Finally, the polaron energy $E_p(R)$ is obtained by combining $E_{st}(R)$ with the strain energy required to establish the potential well which binds the electronic charge carrier:

$$E_p(R) = E_{st}(R) + V_{strain}(R) \equiv \frac{T_e}{R^2} - \frac{V_S}{R^d} - \frac{V_L}{R}. \quad (3)$$

To obtain this result the strain potential energy $V_{strain}(R)$ is recognized as the sum of components from the same regions that contribute $V_S$ and $V_L$. Here, for our simple model, each of these contributions is respectively half of the magnitude of $V_S$ and $V_L$.

This formula for the polaron binding energy in a continuum model is now augmented by a boundary condition. This boundary condition acknowledges that a self-trapped electronic carrier can shrink to no smaller than a minimal structural unit (e.g., ion, atom, bond or molecule). This effect is mimicked by choosing $R$ to be the self-trapped electronic state's radius divided by its value when confined to a single structural unit. Thus our formula only applies for $R \geq 1$.

A minimum of $E_p(R)$ at a finite-radius gives the energy of the strong-coupling polaron in terms of $T_e$, $V_S$ and $V_L$. However, essential features of polaron formation can be elucidated without knowing details. In particular, polaron formation depends critically on the range of the electron-phonon interaction and on the dimensionalities of 1) the electronic-carrier's motion and 2) the medium within which it is immersed.

Figure 1 compares $E_p(R)$ plotted against $R$ for a covalent material ($V_L \rightarrow 0$) with the electronic carrier being a) unconstrained ($d = 3$), b) constrained to a plane ($d = 2$), c) confined to a chain ($d = 1$). For $d = 3$ the only possible finite-radius minimum is at $R = 1$ indicating formation of a small polaron when $V_S > 2T_e/3$. The small polaron is either metastable ($T_e > V_S > 2T_e/3$) or stable ($V_S > T_e$) with respect to the carrier remaining free. Shrinking of a the electronic carrier from $R = \infty$ to $R = 1$ necessitates overcoming the "barrier to self-trapping," $4T_e^3/27V_S^2$. For $d = 2$ the electronic carrier will severely self-trap ($R = 1$) when $V_S > T_e$ thereby forming a small polaron. Alternatively, the carrier will remain free when $V_S < T_e$. For $d = 1$ there is a single finite-radius minimum. For $V_S < 2T_e$ this minimum is at $R = 2T_e/V_S$ corresponding to the formation of a large polaron with energy $-V_S^2/4T_e$. For $V_S > 2T_e$ the minimum is at $R = 1$ corresponding to the formation of a small polaron. Thus, the requirement for stable small-polaron formation becomes more stringent as the electronic dimensionality is decreased: $V_S > T_e$ for $d = 3$ or 2 while $V_S > 2T_e$ for $d = 1$.

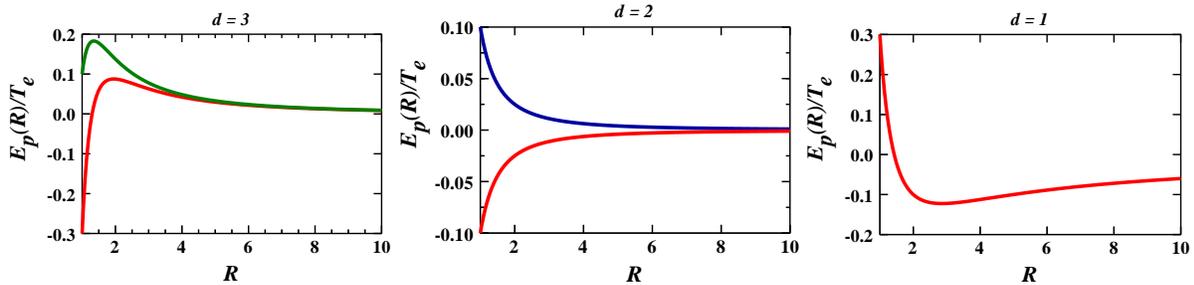

Figure 1 The effect of electronic dimensionality $d$ on adiabatic self-trapping with a pure short-range electron-phonon interaction. Either a metastable or stable small-polaron state coexists with a free-carrier state when $d = 3$. Only a stable small-polaron state or a free-carrier state is permitted when $d = 2$. A stable self-trapped state, illustrated here as a large polaron, always exists when $d = 1$.

Although idealized systems of diminished electronic dimensionality are studied, their realization may be difficult to achieve. In particular, investigations of quasi-one-dimensional systems find that extreme anisotropies are required to achieve the idealized $d = 1$ limit.

Short-range interactions between a carrier and the atoms it contacts are presumably ubiquitous. In this sense, models with only a long-range interaction with displaceable ions are artificial. Moreover, we have already established general features of such polarons. In particular, examination of the above equation for $E_p(R)$ shows that the case of a carrier within a three-dimensional ionic medium without short-range electron-phonon interactions ($V_S = 0$) is isomorphic to the $d = 1$ case with a pure short-range interaction ($V_L = 0$) that was discussed above.

Estimates of $V_S$ obtained from experiments and computations range from over several tenths of an eV to several eV. Meanwhile, $V_L$ is essentially the Coulomb repulsion between two elementary charges confined to a single site, $\approx 10$ eV, multiplied by the factor $[(1/\varepsilon_\infty) - (1/\varepsilon_0)]/2$. Without displaceable ions this factor is near zero. By contrast, for an alkali halide, a prototypical ionic solid, this factor rises to about 0.1.

Combining short-range and long-range components of the electron-phonon interaction can produce several distinct self-trapping scenarios. Consider a three-dimensional model in which the long-range interaction generated by displaceable ions is progressively added to a modest short-range interaction which by itself is presumed to be too weak to produce even a metastable self-trapping, $V_S < 2T_e/3$. Figure 2 plots $E_p(R)/Te$ for progressively increasing values of $V_L$. Curve $a$, for $V_L = 0$. shows no self-trapping. Curve $b$ shows an increase of $V_L$ introducing an energetically stable large-polaron state at $R = 2T_e/V_L$ along with a metastable small-polaron state at $R = 0$. Curve $c$ shows a further increase of $V_L$ stabilizing the small-polaron state with respect to the large-polaron state. Finally, curve $d$ shows the collapse of the large-polaron state into the small-polaron state when $3V_LV_S > T_e^2$.

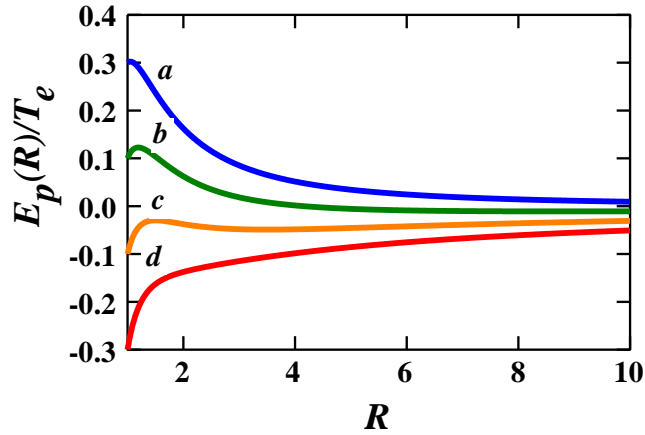

Figure 2 Addition of the long-range component of the electron-phonon interaction facilitates self-trapping. $a$) No self-trapped state with $V_L = 0$. Progressively increasing $V_L$ introduces $b$) metastable small-polaron state with stable large-polaron state and then $c$) stable small-polaron state with metastable large polaron state. $d$) Large-polaron state collapsing yields only a stable small-polaron state.

Atoms generally vibrate about their carrier-induced displaced equilibrium positions. The stiffness constants governing such vibrations are found from the second derivatives of the adiabatic potential with respect to atomic displacements evaluated at their carrier-induced equilibrium position. The stiffness linking positions $\boldsymbol{u}$ and $\boldsymbol{u}'$ is reduced by the polarization of a self-trapped carrier as it adjusts to alterations of its self-trapping potential well induced by atomic vibrations. Thus, a self-trapped carrier is generally associated with both 1) displacements of the equilibrium positions of surrounding atoms and 2) softening of the stiffness constants governing their vibrations. Carrier-induced softening of atoms' vibrations reduces their frequencies and thereby lowers a polaron's free energy. For simplicity, this contribution to polaron stability has heretofore been neglected in our discussion of the formation of strong-coupling polarons. Nonetheless, carrier-induced softening is central to large-polarons' transport.

## Large- and small-bipolaron formation

Charges in solids often pair as singlets. A two-center covalent bond forms as an electron shifts from each of two atoms to being centered between them. Concomitantly, the two atoms move toward one another. The two electrons then overcome their mutual Coulomb repulsion as each of them benefits from its attraction to the bond's two atoms. Similarly, electronic charge carriers can doubly occupy a self-trapping potential well to form a strong-coupling singlet bipolaron.

The model developed to address formation of a strong-coupling polaron can be extended to address the formation of a singlet bipolaron. A bipolaron's two uncorrelated self-trapped electronic carriers each benefits from their doubling of their carrier-induced displacements of atoms' equilibrium positions. Thus, both short-range and long-range contributions to the polaron potential energy are augmented four-fold, $4 = 2 \times 2$. Meanwhile the Coulomb-repulsion between the two self-trapped electronic carriers is reduced by the high-frequency dielectric constant $\varepsilon_\infty$, associated with electronic polarization of the medium. Thus, the energy of a bipolaron $E_{bp}(R)$ becomes:

$$E_{bp}(R) = \frac{2T_e}{R^2} - \frac{4V_S}{R^d} - \frac{4V_L}{R} + \frac{U}{\varepsilon_\infty R} = 2E_p(R) + \frac{U}{\varepsilon_0 R} - \frac{2V_S}{R^d}. \quad (4)$$

The bipolaron energy is then compared with that of two well-separated polarons $2E_p(R)$ where it has been noted that the long-range component of the electron-phonon interaction $V_L$ is simply related to the bare Coulomb interaction between a pair of charges on a minimal structural unit $U$: $V_L = [(1/\varepsilon_\infty) - (1/\varepsilon_0)]U/2$.

The final two terms on the right-hand side of this equation describe the essential physics of bipolaron formation. Bipolaron formation is energetically favorable when the paired carriers' residual mutual Coulomb repulsion, $U/\varepsilon_0 R$, is overwhelmed by the additional binding provided by the short-range electron-phonon interaction, $2V_S/R^d$. In particular, large-bipolaron formation, $R > 1$, is fostered in systems whose electronic carriers have reduced dimensionality.

Bipolaron formation on an embedded plane, $d = 2$, is investigated by comparing the minima of $E_{bp}(R)$ and $2E_p(R)$ with respect to $R$. A planar large-bipolaron will form if the minimum of $E_{bp}(R)$ occurs at $R > 1$ with a lower energy than the minimum value of $2E_p(R)$. In particular, a planar large-bipolaron can form if

$$\frac{4(\varepsilon_0/2\varepsilon_\infty) - 3}{2(\varepsilon_0/2\varepsilon_\infty)^2 - 1} \leq \frac{2V_S}{T_e} \leq 1. \quad (5)$$

Thus, the value of $V_S$ that drives large-bipolaron formation must be large enough to foster its energetic stability without being so large as to collapse the large bipolaron into a small bipolaron. These requirements can be met in unusual circumstances where $\varepsilon_0/2\varepsilon_\infty$ rises above 1.

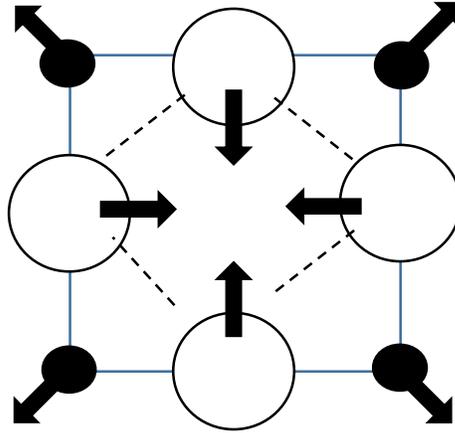

Figure 3 Removing two electrons from four contiguous oxygen anions (large open circles) of the basic structural unit of a $CuO_2$ plane shifts their equilibrium positions inward and those of the copper cations (smaller filled circles) outward.

Cuprates are prime candidates for large-bipolaron formation because 1) their charge carriers are confined to $CuO_2$ planes, $d = 2$, that are 2) embedded within media with exceptionally displaceable ions, $\varepsilon_0 >> 2\varepsilon_\infty >> 1$. An hypothesized microscopic model (Emin, 2017) envisions a large-bipolaron's paired holes in a $CuO_2$ plane of p-doped antiferromagnetic $La_2CuO_4$ occupying out-of-plane non-bonding oxygen orbitals. As depicted in Fig. 3, the core of such a large-bipolaron hole removes two electrons from the four contiguous $O^{2-}$ anions thereby shifting their equilibrium positions inward while shifting the equilibrium positions of the four surrounding Cu cations outward. Concomitantly, electrons transferred from the four oxygen anions to the four copper cations convert each of them from a spin-½ $Cu^{2+}$ to a spin-less $Cu^{1+}$. Such large-bipolarons thereby eliminate the doped material's magnetism. Two electrons then remain distributed over the out-of-plane orbitals of the four oxygen anions: 2 holes + $4O^{2-}$ + $4Cu^{2+}$ → $(O_4)^{2-}$ + $4Cu^{1+}$. The lowest-frequency largest-amplitude optic-mode type radial vibration of these ions has $d$-symmetry.

Boron carbides high-density of low-mobility hopping-type charge carriers appear to pair as singlet small bipolarons since they do not contribute unpaired spins to its magnetic susceptibility (Emin, 2006). Paired carriers' small mutual Coulomb repulsion results from their being constrained to the surface of spheroidal boron-rich twelve-atom icosahedra in a material with a large electronic polarizability, $\varepsilon_\infty >> 1$.

## Disorder-induced small-polaron formation

There is a qualitative dichotomy between states that are only minimally affected by electron-phonon interactions and small-polaronic states. Moreover, rough estimates of the pertinent constants indicate many situations that are close to the boundaries between these two radically different situations.

Disorder, by increasing the localization or slowing of charge carriers, can trigger their collapse into small polarons. Different models have been employed to illustrate these effects.

One approach envisions disorder generating potential wells that confine charge carriers. In particular, disorder is presumed to introduce contributions to the adiabatic potential of the form $-V_{disorder}/R^n$. Electron-phonon interactions then synergistically enhance carriers' localization. As a carrier's disorder-induced localization is progressively increased, it ultimately abruptly collapses into a small-polaron, confining a carrier to a single structural unit. The situation is analogous to that portrayed in Fig. 2, in which a progressively stronger long-range component of the electron-phonon interaction, $-V_L/R$, ultimately drives a carrier's collapse into a small polaron.

An explicit example considers the abrupt collapse of a large-radius donor state in the ferromagnetic semiconductor EuO. The donor electron moves between Eu cations' $S = 7/2$ spins. The intra-atomic exchange $I$ between the donor electron and each of the Eu cations it contacts fosters their alignment even as rising temperature drives misalignment of the spins of other Eu cations. As the temperature is raised toward the EuO Curie temperature, $T_C \approx 70$ K, the large-radius donor abruptly collapses to a small-radius donor. Concomitantly, the number of aligned Eu spins shrinks. The product of the temperature and the difference between the spin-misalignment entropy associated with large-radius and small-radius states is the transition's latent heat. The free-energy functional for this system is:

$$F_{bmp}(R,T) = \frac{\left[T_e - C\left(\frac{I}{kT_C}\right)kT\right]}{R^2} - \frac{V_S}{R^d} - \frac{(V_L + V_d)}{R}, \quad (6)$$

where C denotes a numerical constant which for EuO is estimated to be between 0.01 and 0.1 and $V_d$ represents the electron's Coulomb attraction to its donor (e.g. a dopant or an oxygen vacancy). The minimum of this free-energy functional can collapse from that for a large-radius donor, $R \gg 1$, to that for a small-polaronic state, $R = 1$, as the temperature is raised above 50 K.

Strikingly, the magnetic disorder which drives this transition is controlled by the temperature and is therefore readily reversible. This collapse also can be reversed by applying a magnetic field which fosters alignment of the Eu cations' spins. In particular, the temperature of the collapse rises as the square-root of the applied magnetic field.

Disorder can also induce the self-trapping of freely moving electronic excitations. For example, excitons in crystalline TlCl or TlBr move freely while those in the mixed crystal TlCl$_x$Br$_{1-x}$ self-trap. Holes injected into crystalline orthorhombic sulfur appear to move with a moderate mobility which falls with increasing temperature until the crystal's softening temperature is reached. Then, once the crystal's S$_8$ molecules lose registry with one another, the mobility drops precipitously. As the temperature is raised further the mobility rises in an Arrhenius manner, typical of small polarons. Charge carriers in crystalline Si and Ge have large Hall mobilities that fall with increasing temperature. By contrast, the Hall mobilities of amorphous Si and Ge are very low, rise in an Arrhenius manner with increasing temperature and display the sign anomalies expected for both electron and hole small polarons.

Nearest-neighbor inter-atomic separation are nearly the same for crystalline and amorphous semiconductors. By contrast, amorphous covalent semiconductors have large ($> 10°$) variations of their bond angles. The electronic transfer energies of electrons in anti-bonding states and holes in bonding states are very sensitive to bond-angle variations.

The effect of transfer-energy variations on small-polaron formation is now considered. Self-trapping was previously addressed by determining whether an electronic carrier will be bound within the potential well generated by fixing surrounding atoms at their carrier-induced displaced equilibrium positions (the adiabatic limit). With just the ubiquitous short-range component of the electron-phonon interaction, an electronic carrier in a three-dimensional deformable continuum will either self-trap as a small-polaron or remain free.

An electronic carrier remains free if its energy band extends below the small-polaron energy. Disorder-induced transfer-energy variations by themselves can then introduce sites at which small-polaron formation is stable. Moreover, transfer energies emanating from such sites are suppressed since moving a self-trapped electronic carrier requires substantial movement of the associated atoms. Sites with stable small-polaron formation are thereby extended to some adjacent sites. Ultimately, through this nonlinear feedback mechanism, modest transfer-energy disorder can trigger global small-polaron formation. All told, transfer-energy disorder significantly reduces the electron-phonon coupling strength required to stabilize small-polaron states with respect to free-carrier states.

Such a small-polaron collapse requires transfer-energy variations (non-diagonal disorder) that are comparable to the small-polaron energy and the electronic carrier's bandwidth. By contrast, severe Anderson localization requires diagonal-disorder energies that are an order of magnitude greater than the electronic carrier's bandwidth, as envisioned for impurity conduction in very lightly doped compensated semiconductors.

## Coherent and incoherent polaron transport

A strong-coupling polaron's motion requires shifting surrounding atoms away from the displaced equilibrium positions they assume when generating the potential well which binds its self-trapped electronic carrier. A self-trapped electronic carrier only moves in response to these atoms' movements.

In the most extreme case, atoms may move enough to completely eliminate the self-trapping potential well thereby liberating its electronic carrier. The electronic carrier can then move away until atoms elsewhere in the material bind it within another self-trapping potential well. This transport mode corresponds to free-carrier motion limited by occasional self-trapping.

Transport of a strong-coupling polaron occurs when the centroid of its self-trapped electronic carrier moves between sites in response to atoms' movements shifting the potential well that binds it. The motion of a strong-coupling polaron is regarded as coherent when its self-trapped electronic carrier extends over multiple sites. As such, the self-trapped electronic carrier of a large polaron moves continuously in response to the classical motion of surrounding atoms. In particular, the mass for a three-dimensional large-polaron is $\approx E_p/(\omega R_p)^2$, where $E_p$ and $R_p$ respectively denote the polaron energy and radius while $\omega$ represents the frequency characterizing vibrations of the relevant atoms. The effective mass for a one-dimensional large-polaron interacting with acoustic phonons is $4E_p/c_s^2$, where $c_s$ denotes the sound velocity. These polaron masses are generally orders of magnitude greater than the free-electron mass.

Since the huge masses of strong-coupling large polarons result from the requisite atomic motion, they can only be observed at frequencies below those characterizing atomic motion. By contrast, masses measured at high frequencies correspond to those of electronic charge carriers, e.g., an electronic carrier moving within the self-trapping potential well that binds it.

Distinctively, the transport of a small polaron is usually incoherent. Its severely localized self-trapped electronic carrier moves discontinuously between sites in response to atomic movements. Small-polaron transport is generally described as occurring via sequences of phonon-assisted hops.

It should be noted that a polaron in a crystal can equally well occupy any of its geometrically equivalent sites. Therefore, its eigenstates extend throughout the crystal. However, unlike purely electronic Bloch states, polaron eigenstates transfer the pattern of displaced atomic equilibrium positions collaterally with the self-trapped electronic carrier. Therefore, the width of such a polaron band is proportional to the product of the overlaps between each atom's vibrational wavefunction when its self-trapped electronic carrier is shifted between neighboring sites. For example, the maximum nearest-neighbor transfer energy associated with a small-polaron adiabatically shifted between adjacent sites with the electron-phonon interaction taken as purely short-range is

$$t_{sp} \cong \sqrt{\frac{E_p \hbar \omega}{\pi}} e^{-\left(\frac{E_p}{\hbar \omega}\right) coth\left(\frac{\hbar \omega}{2kT}\right)}, \quad (7)$$

where the strong-coupling condition insures that $E_p >> \hbar \omega$. It should be observed that $t_{sp}$ is 1) very small, $t_{sp} << \hbar \omega$, 2) decreases with rising temperature, and 3) vanishes in the classical limit, $\hbar \to 0$. The extreme narrowness of a small-polaron band implies that its coherent motion associated with it will be suppressed by any realistic perturbations.

## Large-polaron transport

The mobility $\mu$ of a coherently moving carrier of charge $q$ can be schematically written in terms of the average of the product of its velocity $v$ and its mean-free-path $\ell$: $\mu = e<v\ell>/kT$. The minimum mobility for coherent transport corresponds to the carrier mean-free-path falling to its size, its de Broglie wavelength $\lambda = h/p = h/m^*v$, where $m^*$ denotes the carrier's effective mass: $\mu_{min} = (eh/m^*kT)$. The minimum coherent-transport mobility associated with a carrier mass equal to the free-electron mass is about 300 cm$^2$/V-sec at room temperature. Since this minimum mobility is inversely proportional to the carrier effective mass, room-temperature mobilities that are much lower than this minimum imply very large effective masses such as those of large polarons.

A large-polaron's transport depends critically on the phonon modes involved in constructing the potential well in which its electronic carrier is self-trapped. These modes are locally softened by relaxation of the large-polaron's self-trapped electronic carrier to their atoms' vibratory displacements. A large polaron's local softening of these vibrational modes impedes passage of the solid's indigenous phonons through it. This regional phonon softening is the origin of the scattering between a large polaron and ambient phonons.

The huge mass of a strong-coupling large polaron insures that its thermal momentum near room temperature exceeds that of the phonons with which it scatters. Thus, while conventional electronic charge carriers undergo wide-angle scattering by phonons, a large polaron primarily "reflects" incident phonons. This difference may manifest itself in a solid's carrier-induced thermal conductivity.

The dynamics of the scattering between a large polaron and a solid's phonons depends on the velocity with which vibrational energy can be transported through a solid. This feature of phonon transport is governed by their vibrational dispersion. Two examples illustrate contrasting limits. Acoustic phonons have substantial dispersion with the velocity of their longitudinal branch being just that of sound, $c_s$. Optic phonons are frequently represented as Einstein oscillators, which possess zero vibrational dispersion. Therefore, the huge mass of a large polaron causes it to move more slowly than the acoustic phonons which it scatters. However, a large polaron moves faster than Einstein oscillators that are static.

The scattering rate for a large polaron by acoustic phonons is

$$\frac{1}{\tau_{lp}} \cong \left\langle N_{ac} \sigma \left(\frac{2\hbar q}{m_{lp}}\right) \right\rangle \cong \frac{kT}{m_{lp} c_s R_p}, \quad (8)$$

where $N_{ac}$ denotes the density of thermally available acoustic phonons, $\sigma$ represents the cross-section for scattering between a large polaron and acoustic phonons, and the final factor is the change of a large polaron's velocity generated by its reflection of an acoustic phonon of momentum $\hbar q$. The final expression on the right results from evaluating the average at or near room-temperature upon recognizing that the relevant phonons have wavelengths comparable to the polaron radius $R_p$. It is noteworthy that this scattering rate is inversely proportional to the large-polaron mass, $m_{lp}$. That is, a large polaron is not easily scattered by acoustic phonons which possess much smaller momenta.

A distinctive characteristic of a large polaron is its extremely long scattering time, $\tau_{lp} \approx (E_p/kT)/\omega \gg 1/\omega$. This feature manifests itself by the real part of the large-polaron conductivity as a function of applied frequency $\Omega$, proportional to $1/(1 + \Omega^2 \tau_{lp}^2)$, only being appreciable at frequencies well below the phonon frequency $\omega$. Moreover, as will be discussed subsequently in the section on polaron's optical properties, a large polaron's conductivity also has a very broad peak beginning at frequencies well above that of its characteristic phonon. Thus, there is a pseudo gap between the low-frequency, $\Omega \ll \omega$, and the high frequency, $\Omega \gg \omega$, contributions to a large polaron's conductivity and related absorption. By contrast, a conventional electronic carrier's scattering time is very long, $\tau \gg \omega$. Thus, the frequency dependence of the real part of a conventional electronic carriers' conductivity, its Drude conductivity, is appreciable at frequencies that are many orders of magnitude higher than that for a large polaron.

The large-polaron mobility corresponding to its long scattering time $\tau_{lp}$ and huge effective mass $m_{lp}$ is:

$$\mu_{lp} \equiv e \frac{\tau_{lp}}{m_{lp}} = \frac{e c_s R_{lp}}{kT}. \quad (9)$$

This mobility is typically comparable to 1 cm$^2$/V-sec at room temperature, much smaller than that for conventional electronic charge carriers.

This mobility's monotonic decrease with increasing temperature distinguishes it from that for a large polaron interacting with dispersion-less phonons, Einstein oscillators. The mobility arising from the scattering of a large polaron with Einstein optic vibrations involves a competition between two factors. First, the density of optic phonons increases with rising temperature. Second, the efficacy of their scattering of a large-polaron diminishes as its de Broglie wavelength decreases with increasing temperature. Strikingly, this effect is reported to make a large-polaron's mobility rise with increasing temperature above a small fraction of the optic-phonon temperature.

## Small-polaron transport: Elemental hops

A small-polaron's severe localization renders its transport incoherent. This incoherent transport is best described as proceeding via successions of thermally assisted jumps. A small-polaron's hop is governed by movements of the atoms to which its self-trapped electronic carrier is coupled. The motions of these atoms are controlled by the system's adiabatic potential.

The processes governing an elemental small-polaron hop depends on its two-site adiabatic potential. A simple model has the portion of the adiabatic potential governing a jump depending on a single relative atomic configurational coordinate. A small-polaron hop then corresponds to motion of a fictitious particle with an appropriate reduced atomic mass moving between minima of this two-site adiabatic potential.

Figure 4 shows the two-site adiabatic potential $V_{ad,2}(x)$ as a function of the relative atomic-configuration coordinate $x$. An energy barrier exists between equivalent minima located at $x_i$ and $x_f$. A fictitious particle's occupation of one of these two equivalent minima corresponds to the self-trapped carrier occupying one of the two equivalent sites between which this small polaron can move.

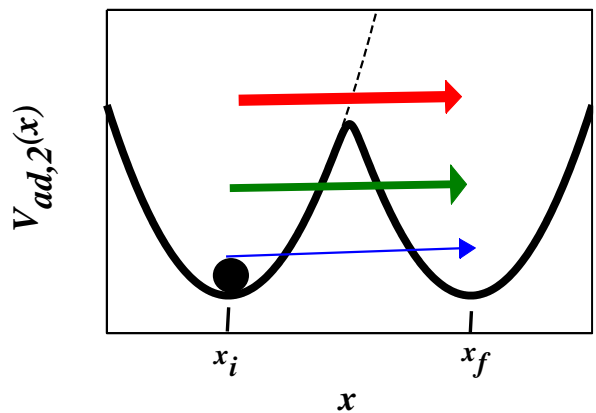

Figure 4 The two-site adiabatic potential energy for a small-polaron hop is plotted against the relative atomic configurational coordinate $x$. The dashed line indicates the extension of the adiabatic potential for occupation of the initial site in the absence of electronic transfer to the hop's final site.

At low temperatures the charge carrier moves from initial to final sites as the associated atoms undergo quantum-mechanical tunneling between the configurations they assume when the electronic carrier occupies corresponding initial and final sites. This relatively slow atomic tunneling process is depicted in Fig. 4 by the thin blue arrow passing through the thick portion of its energy barrier.

As the temperature is raised, the amplitudes of atoms' oscillations increase. The fictitious particle, the large black dot in Fig. 4, then rises higher in its potential well before tunneling through a relatively thin portion of this figure's energy barrier. The green arrow depicting this tunneling is thicker than that of the blue arrow to indicate that the tunneling rate increases as the width of the energy barrier decreases. The decrease in the energy barrier indicates reduced disparity of the positions of atoms involved in their quantum-mechanical tunneling.

At sufficiently high temperatures, the fictitious particle can move between potential wells by transcending the energy barrier rather than tunneling through it. The predominant small-polaron hopping process then does not involve atoms' tunneling. As indicated by the thickness of the red arrow the small-polaron jump rates exceeds that of processes requiring atomic tunneling.

Figure 5 plots the elemental adiabatic small-polaron jump rate for a hop between equivalent sites. At high temperatures the jump rate is Arrhenius. Its activation energy corresponds to the energy difference between the minima of the adiabatic potential and the height of the energy barrier shown in Fig. 4. This activation energy is the minimum energy required to displace atoms from the equilibrium positions they assume when a hop's initial site is occupied so as to bring the electronic energies of the jump's initial and final sites into coincidence. As the temperature is lowered the small-polaron jump rate becomes non-Arrhenius. This behavior indicates the progressive freeze out of multi-phonon processes requiring the absorption of more energy than is the minimum consistent with the requirement of energy conservation. The details of these non-Arrhenius temperature dependences depend upon the particular phonons that are coupled to the electronic charge carrier. However, the transition between Arrhenius and non-Arrhenius behavior typically occurs at about one third of the characteristic phonon temperature.

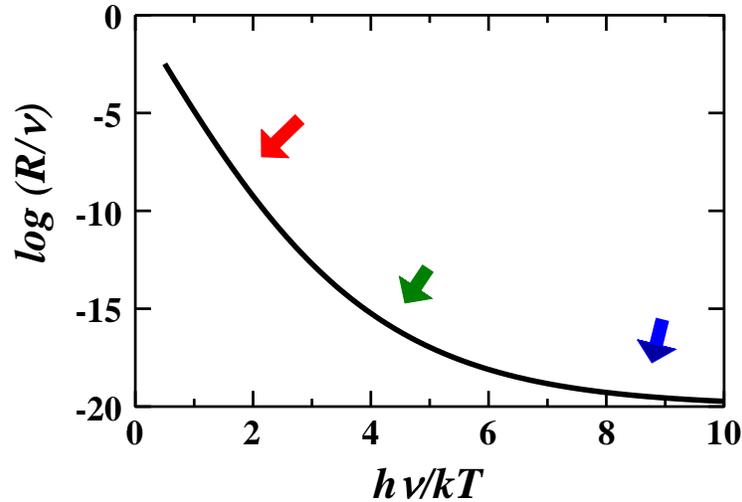

Figure 5 The rate for an elemental adiabatic small-polaron jump between equivalent states divided by the hop's characteristic phonon frequency $\nu$ is plotted versus $h\nu/kT$. The colored arrows correspond to those in Fig. 4 which indicate the processes which predominate in each temperature regime.

The example presented here presumed that the sites between which hopping occurs are equivalent to one another, as they are in a perfect crystal. Nonetheless, except at very low temperatures, the adiabatic small-polaron jump rate between inequivalent sites is usually only modified by the factor $\exp(-\Delta/2kT)$, where $\Delta$ denotes the difference between final and initial sites. At extremely low temperatures the energy absorbed in a hop falls to the minimum energy consistent with energy conservation. Then the rate for a hop upward in energy is proportional to $\exp(-|\Delta|/kT)$ while that for a hop downward in energy is temperature independent.

A small-polaron's adiabatic jump rate's temperature dependence (e.g., its activation energy) implicitly measures the distances that atoms surrounding a jump's pair of sites must move to generate a hop. This effect depends on the range of an electronic charge carrier's electron-phonon interaction. Nonetheless, the temperature-dependence of a small-polaron's adiabatic jump rate generally increases as the separation between initial and final sites is increased. Thus, the activation energy for a hop increases with its distance. By itself, this effect favors short hops over longer hops.

Application of a dc electric field $E$ alters the temperature dependences of the jump rates for hops emanating from a site. This effect is greatest for long jumps. The resulting effect usually increases the net jump rate from a site. Frenkel-Poole behavior denotes the applied electric-field $E$ increasing the net jump rate from a site by the factor $\exp[CE^{1/2}/kT]$, where $C$ denotes a constant.

Arrhenius small-polaron hopping is adiabatic when its self-trapped electronic carrier readily moves between sites in response to atoms' movements establishing the requisite coincidence event. The magnitude of the electronic transfer energy $t_{tr}$ must then exceed the energy spread $\Delta E_{dur} \approx [\hbar\omega(E_A kT)^{1/2}]^{1/2}$, determined by the Heisenberg uncertainty principle, associated with the establishment of an energy coincidence. Figure 6 displays the probability $P$ of a hop being adiabatic plotted against the ratio $t_{tr}/\Delta E_{dur}$.

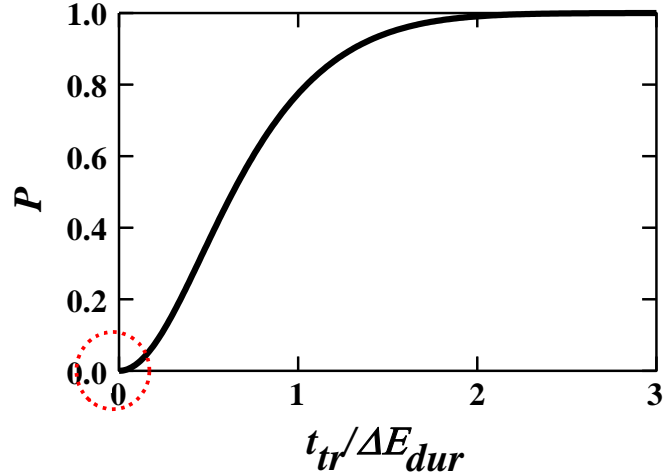

Figure 6 The probability $P$ of a self-trapped electronic carrier adiabatically moving between sites during a coincidence event is plotted versus $t_{tr}/\Delta E_{dur}$. The circle indicates the non-adiabatic regime where $P \propto t_{tr}{}^2$.

For computational simplicity, initial studies of small-polaron hopping presumed that the inter-site transfer energy of an electronic carrier is so small that it rarely moves between sites even when atomic movements enable it to do so, $P \approx 2(t_{tr}/\Delta E_{dur})^2 << 1$. However, estimates of $t_{tr}/\Delta E_{dur}$ often imply adiabatic hopping, $P \approx 1$. Fortunately, the value of $P$ can often be estimated from experiments. In particular, the pre-exponential factor for Arrhenius small-polaron hopping mobility is $(ea^2\nu/kT)P$, where $a$ denotes the separation between small-polaron sites. In addition, the corresponding pre-exponential factor of the dc conductivity for thermally generated small polarons is $(e^2\nu/akT)P$. Typical estimates of the bracketed expressions in these two pre-exponential factors are respectively about 1 cm²/V-sec and $10^3$ S/cm. The frequent observations of pre-exponential factors that are close to these values implies that their small-polaron hopping is adiabatic, $P \approx 1$.

Small bipolarons also move by thermally assisted hopping. At low temperatures small-bipolarons move together as singlet pairs. However, at high temperatures pairs separate and charge transport becomes predominantly that of small polarons. The activation energy for pair-breaking and subsequent small-polaron hopping is the sum of two contributions. The contribution from pair separation is half of the binding energy of the bipolaron with respect to forming two separate polarons, $(2E_p - U)/2$, where $U$ represents a pair's on-site Coulomb repulsion. As before, the activation energy for small-polaron hopping is below $E_p/2 - t_{tr}$. Therefore the activation energy for breaking a pair and subsequent small-polaron hopping is less than $3E_p/2 - U/2 - t_{tr}$. By comparison, the activation energy for the pair jumping as a unit is $2E_p - t_{2tr}$, where $t_{2tr}$ denotes the transfer energy for the paired electronic carriers moving as a unit in response to a coincidence event, $t_{2tr} << t_{tr}$. Thus, due to its smaller activation energy, high-temperature small-bipolaron transport is dominated by pair breaking. Distinctively, the paramagnetic susceptibility then garners a high-temperature Arrhenius contribution from the collateral generation of unpaired self-trapped electronic charge carriers. This phenomenon appears to occur in boron carbides, $B_{12+x}C_{3-x}$ with $0.15 < x < 1.7$, where $x/2$ denotes the fraction of unit cells with hole-like small bipolarons (Emin, 2006).

Small-polaron hopping is qualitatively different from very low temperature inter-impurity hopping conduction in lightly doped compensated crystalline semiconductors (e.g., P and B doped Si). The small doping concentrations (e.g., $10^{15}$ cm⁻³) imply very long (e.g., 1000 Å) non-adiabatic jumps. The large dopant radii (e.g., 20 Å and 45 Å in Si and Ge, respectively) and small conductivity activation energies (e.g., $10^{-4} - 10^{-3}$ eV) imply each hop involves absorption or emission of only a single very long wavelength acoustic phonon. As such, the elemental jump rates are proportional to $\exp[-(\Delta + |\Delta|)/2kT]$ in the very low-temperature limit and proportional to $T$ at higher temperatures. All told, inter-impurity hops are primarily non-adiabatic, proportional to $t_{tr}^2$, and between sites whose near-random energies are determined by their proximity to the doped-semiconductors' charged impurity centers. Rates for individual inter-impurity hops therefore involve competitions between 1) transfer energies which favor short jumps and 2) energy absorptions which favor long jumps. By itself, these effects would lead to variable-range hopping. By contrast, small-polaron hops are usually short (< 10 Å), adiabatic, multi-phonon. As such, small-polaron jump rates are non-Arrhenius below a fraction of the phonon temperature and Arrhenius at higher temperatures with activation energies between 0.1 and 1 eV.

## Small-polaron relaxation phenomena

Small-polaron hopping involves local accumulation and dissipation of atoms' vibrational energy. In particular, the elemental thermally assisted small-polaron hop described in Fig. 4. begins as the amplitudes of atomic vibrations increase while the self-trapped electronic carrier occupies its initial site and ends with vibrational energy being dissipated to surrounding atoms after the electronic carrier transfers to its final site. Thus, the jump process involves vibrational dispersion.

Early studies of small-polaron hopping only explicitly addressed the actual inter-site transfer of an electronic charge carrier. The processes governing local accrual and subsequent dispersal of vibrational energy were ignored. Rather, the energies of local vibrations were assumed to retain their equilibrium distribution. This procedure is justified in the non-adiabatic limit since the probability of an electronic transfer in response to atoms assuming favorable configurations is then arbitrarily small. As such, dispersion-related transfers of atoms' vibrational energies are always fast enough to establish equilibrium prior to an electronic charge carrier transferring between sites. However, as noted earlier, small-polaron hopping is typically adiabatic. In this situation, the inter-site transfer of an electronic charge carrier occurs in less than a vibrational period while the accumulation and dissipation of vibrational energy is much slower. Transport of vibrational energy then affects small-polaron hopping.

Following a hop, vibrational energy is dissipated away from the involved sites. In a three-dimensional vibrational system, the time for the sites' vibrational energy to fall to $kT/2$ following a hop with activation energy $E_A$ is $\tau_{\text{relax}} \approx (1/\Delta\omega)(2E_A/kT)^{2/3}$, where $\Delta\omega$ denotes the spread of frequencies characterizing the phonons involved in the jump. This relaxation time will generally exceed that governing atoms' vibrations: $\omega\tau_{\text{relax}} >> 1$, where $\omega >> \Delta\omega$ and $E_A >> kT$. Thus, relaxation from one hop may not be completed before vibrations establish another adiabatic jump. Hopping will then be correlated. Adiabatic small-polaron hops occur in flurries separated by dormant intervals. Within a flurry, the probability of back and forth hopping between a pair of sites is greatly enhanced while jumps to previously unvisited sites are also enhanced but to a lesser extent.

As shown in Fig. 4, a self-trapped electronic carrier's adiabatic over-the-barrier shuttling between a pair of sites is governed by the broad double-well adiabatic potential. By contrast, a carrier confined to the initial site is governed by the single harmonic potential depicted by the dashed curve. The difference in vibrational free energy between that with the carrier transferring back-and-forth and that with it confined to its initial site, $\Delta F_{vib} < 0$, augments the jump rate. The jump rate is also halved because the shuttling carrier only ends on the final site half of the time. The formula for the Arrhenius jump rate then becomes:

$$R_{M-N} = \left[\frac{\nu}{2} exp\left(-\frac{\Delta F_{vib}}{kT}\right)\right] exp\left(-\frac{E_A}{kT}\right) \rightarrow \left[\frac{\nu}{2} exp\left(\frac{E_A}{t_{tr}}\right)\right] exp\left(-\frac{E_A}{kT}\right), \quad (10)$$

where the final expression applies when $t_{tr}^2 > E_A kT$ (Emin, 2013). In accord with the observed Meyer-Neldel compensation effect, this expression has its temperature-independent factor increasing exponentially with the hopping activation energy $E_A$. The rise of the jump rate's temperature-independent factor with increasing $E_A$ partially compensates for the decrease of the rate's thermally activated factor.

Electronic charge carriers injected into condensed matter must dissipate a significant amount of energy in the process of their becoming self-trapped. In particular, for an injected electronic carrier to form an equilibrated small polaron an energy at least comparable to its binding energy (greater than several tenths of an eV) must be transferred into the vibrations of associated atoms. The resulting very large athermal vibrations can facilitate establishing the energetic coincidences required for small-polaron hopping. Thus, transient hopping can occur without significant additional absorption of vibrational energy. As a result, transitory adiabatic small-polaron hopping will be nearly activation-less. It will occur with a mobility of about $ea^2 v/kT \approx 1$ cm$^2$/V-sec. Time-of-flight measurements report such transient mobilities in materials whose steady-state thermally activated mobilities suggest small-polarons.

## Polarons' Seebeck coefficients

The Seebeck coefficient $S$ measures the open circuit e.m.f. $\Delta\phi$ developed across a material in response to the application of the temperature difference $\Delta T$: $S \equiv \Delta\phi/\Delta T$. In physical terms the Seebeck coefficient $S$ is the entropy transported with a carrier divided by its charge $q$. A charge carrier's Seebeck coefficient is related to its Peltier heat, $\Pi = qTS$, the heat transported with a charge carrier.

Analysis of Seebeck coefficient measurements aids understanding charge transport generally and polaron transport, in particular. Combined measurements of the Seebeck coefficient and the dc conductivity distinguish between small-polaron hopping and other models of hopping transport. Large and unusual polaron-based Seebeck coefficients have been exploited to produce novel thermoelectric devices.

There are several different contributions to the Seebeck coefficient. The primary contribution is often the change of the entropy-of-mixing upon adding a carrier of energy $E$ and charge $q$ to a system of chemical potential $\mu$: $<E - \mu>_\sigma/qT$, where the brackets indicate an average of states' energies weighted by their contributions to the electrical conductivity $\sigma$.

It is sometimes useful to consider the entropy-of-mixing contributions to Seebeck coefficients as functions of $c$, the ratio of the number of charge carriers to the number of a material's equivalent structural units. When carriers exist in an energy band that is narrower than $kT$ the Seebeck coefficient becomes temperature independent equal to $(k/e)\ln[(2 - c)/c]$ for fermions that can doubly occupy a state, $(k/e)\ln[2(1 - c)/c]$ for fermions that can singly occupy a state but with a two-fold spin degeneracy, and $(k/2e)\ln[(2 - c)/c]$ for carriers that pair as singlet bipolarons which can only singly occupy a state.

The magnitude of the entropy-of-mixing contribution to a semiconductor's Seebeck coefficient is usually much greater than $(k/e) = 86$ μV/K. Nonetheless, the third law of thermodynamics requires that the entropy and hence the Seebeck coefficient vanish at absolute zero. The magnitude of a semiconductor's Seebeck coefficient drops precipitously to near zero when the ratio of $kT$ to the carrier bandwidth becomes very small (Emin, 2016). Thus, the temperature of this dramatic fall indicates whether the carrier's energy band is consistent with conventional charge carriers, several eV, or with polarons, less than the phonon energy, $0.03 - 0.1$ eV.

The magnetic moment of a charge carrier is found by measuring the magnetic-field dependence of the Seebeck coefficient. In particular, the magnetic field $H$ breaks the two-fold spin degeneracy of singly occupied states. The Seebeck coefficient's spin degeneracy factor then becomes $(k/e)\{\ln[2\cosh(\mu_B H/kT)] - (\mu_B H/kT)\tanh(\mu_B H/kT)\}$, where $\mu_B$ represents the carrier's magnetic moment. A significant $H$-dependent fall of this factor from $(k/e)\ln(2)$ to zero is expected below 10 K at fields exceeding 10 T. The absence of this Seebeck-coefficient magnetic-field dependence implies that charge carriers have no net magnetic moment since they have paired into singlet bipolarons.

The conductivity activation energy for Arrhenius small-polaron hopping between equivalent states is the sum of that for carrier generation and that of the mobility: $(E_S - \mu) + E_A$, while the characteristic energy of the Seebeck coefficient is simply $E_S - \mu$. Therefore, the difference of the characteristic energies obtained from dc conductivity and Seebeck coefficient measurements is just that of the mobility $E_A$. Observing a significant value of $E_A$ is often taken as evidence of small-polaron hopping.

As in a semiconductor's band tail, phonon-assisted hopping can be envisioned as occurring between severely localized states that are strongly coupled to phonons and weakly-localized states that are weakly coupled to phonons. The transported vibrational energy from a hop between states $i$ and $j$ then contributes to the Seebeck coefficient:

$$S_{i,j} = \frac{E_i \left( \frac{\Gamma_j}{\Gamma_i + \Gamma_j} \right) + E_j \left( \frac{\Gamma_i}{\Gamma_i + \Gamma_j} \right) - \mu}{qT}, \quad (11)$$

where and $\Gamma_i$ and $\Gamma_j$ denotes the electron-phonon coupling of these states. When one initial state is much more strongly coupled to phonons than the other state, $\Gamma_s >> \Gamma_w$, the Seebeck coefficient between them is always related to the more weakly coupled state, $(E_w - \mu)/qT$. Thus, the Seebeck energy for thermally exciting carriers from deep within a bandgap to near a band- or mobility-edge is comparable to that of the conductivity. By contrast, the Seebeck energy for small-polaron hopping is significantly smaller than the conductivity activation energy.

Self-trapped electronic carriers adjust to displacements of surrounding atoms from their equilibrium positions. These adjustments generally reduce the frequencies of these atoms' vibrations. The associated carrier-induced increase of the net vibrational entropy contributes to the Seebeck coefficient. With increasing temperature this contribution to the Seebeck coefficient rises from zero at absolute zero to its high-temperature limit: $(k/q)\Sigma_i (-\Delta\omega_i/\omega_i)$ for moderate frequency shifts, $(-\Delta\omega_i/\omega_i) << 1$, where $\Delta\omega_i$ denotes the carrier-induced change of the frequency of the $i$-th vibration mode $\omega_i$.

By itself, carrier-induced softening also fosters vibrational energy being transported with a phonon-assisted hop. For example, the contribution to the Seebeck coefficient from a high-temperature Arrhenius small-polaron hop between equivalent states is $(1/qT)\Sigma_i (-\Delta\omega_i/\omega_i)(E_{p,i}/2)$, where $E_{p,i}$ denotes the $i$-th vibrational mode's contribution to the small-polaron's binding energy. This contribution rises with decreasing temperature to a peak at a fraction of the characteristic phonon temperature and then falls to zero at absolute zero. The large Seebeck coefficients, $> 200$ μV/K above 300 K, seen in boron carbides despite their very high bipolaron densities are attributed to bipolaron-induced vibrational softening (Emin, 2006).

Carrier-induced magnetic changes also manifest themselves in the Seebeck coefficient. In particular, the previously-suggested model of large-bipolaron formation within the $CuO_2$ planes of doped-$La_2CuO_4$ has each large-bipolaron removing a spin-1/2 from each of its four adjacent Cu cations. Distinctively, this effect lowers the Seebeck coefficient of such a large bipolaron by eliminating the entropy associated with these spins' orientations. In the high-temperature paramagnetic region the spin-related reduction of these large-bipolarons' Seebeck coefficients reaches $-(k/e)[4\ln(2)]$.

## Polarons' Hall mobilities

A current produced by an electric field tends to be deflected by the application of a transverse magnetic field. This deflection angle, the Hall angle, divided by the magnetic field's strength defines the Hall mobility $\mu_H$. By contrast, a charge carrier's drift velocity divided by the electric field driving its flow defines the usual mobility $\mu$. Since $\mu_H$ and $\mu$ measure different phenomena, they generally differ from one another. Nonetheless, a simple treatment of coherent transport in an energy band that is much wider than the thermal energy $kT$ yields $\mu_H = \mu$. However, significant differences between $\mu_H$ and $\mu$ emerge for coherent charge transport in energy bands that are narrower than the thermal energy $kT$. The dissimilarity between these two mobilities becomes profound for charge carriers that move incoherently by thermally assisted hopping. Then $\mu_H$ and $\mu$ differ from one another in magnitude, temperature dependence and often even in sign. These distinctive features facilitate identifying charge carriers as small polarons.

The Hall Effect centers on the deflection of charge carriers in a magnetic field. As such, the small-polaron Hall Effect requires considering at least a planar array of sites. Models for studying the Hall Effect envision a magnetic field applied perpendicular to a planar lattice comprising tiles of equilateral triangles and a planar lattice comprising square tiles.

A magnetic field alters transport among the sites of a planar lattice's primary structural unit, equilateral triangle or square. The magnetic field contributes a site-dependent phase factor to each site's electronic wavefunction. Electronic transfer energies therefore garner magnetic phase factors that depend on the linked sites. From this starting point two complementary approaches have been employed to calculate the small-polaron Hall Effect in the high-temperature Arrhenius regime.

The non-adiabatic limit presumes very small electronic inter-site transfer energies. Then an electronic carrier rarely moves between sites in response to opportunities presented by atoms' motions. Concomitantly, this approach's basis states are each localized on a site. The magnetic field's influence then results from interference between different transfer processes. For example, interference on a triangle occurs between a direct inter-site transfer and an indirect transfer through the third site. This interference depends on the magnetic flux through a lattice's primary structural unit.

The adiabatic limit presumes that an electronic carrier always responds to atomic movements by moving to the lowest-energy site of its molecular orbital. This behavior can be portrayed with a generalization of the one-dimensional configuration-coordinate diagram of Fig. 4. In particular, an electronic charge carrier moving amongst the three sites of an equilateral triangle corresponds to a fictitious particle moving amongst three mutually adjacent quasi-harmonic potential wells of a two-dimensional adiabatic potential surface. The magnetic field converts the real wavefunction describing the electronic carrier's non-degenerate molecular orbital into a complex wavefunction. As a result, terms proportional to atomic momenta of the adiabatic Hamiltonian survive to act like a fictitious vector potential. The magnetic field corresponding to this fictitious vector potential is proportional to the real applied magnetic field. This magnetic field affects the motion of the adiabatic potential's fictitious particle. Moreover, the fictitious magnetic field is strongest along the ridges separating adjacent quasi-harmonic potential wells. The magnetic field thereby shifts fictitious-particle trajectories corresponding to shifting the final site of an electronic carrier's hop. Analysis of these trajectories yields the adiabatic small-polaron Hall Effect.

The Hall mobility is ideally intrinsic, unaffected by trapping, since the Lorentz force only deflects a moving charge carrier. The magnitude of the small-polaron Hall mobility is at most comparable to 1 $cm^2$/V-s. Most importantly, the activation energy of the small-polaron Hall mobility in its high-temperature Arrhenius regime is always less than that of the conventional mobility. In particular, although the ratio of these two activation energies depends on the details of the model, the activation energy of $\mu_H$ for small-polaron hopping is often about one-third that of $\mu$. Furthermore, the Hall mobility activation energy even can be small enough that the temperature dependence of $\mu_H$ is dominated by that of its pre-exponential factor. In these instances, the Hall mobility will decrease with increasing temperature.

The sign of the Hall coefficient in wide-band conventional semiconductors is 1) negative for electrons in states with positive effective mass near conduction-band minima and 2) positive for vacant electronic states with negative effective mass near valence-band maxima. This simple situation does not always apply for energy bands which are narrower than the thermal energy $kT$. Moreover, Hall-Effect sign anomalies, opposite those for charge carriers in conventional semiconductors, are common for small-polaron hopping.

The sign of the Hall coefficient for hopping conduction is given by

$$sgn(R_H) = sgn\left[(-q)(\delta)^{n+1}\prod_{i=1}^{n} t_{i,i+1}\right], \quad (12)$$

where $q$ denotes the bona-fide carrier's charge (negative for electrons). In addition, $\delta$ indicates the filling of the states between which carriers hop ($\delta = -1$ for nearly empty states and $\delta = 1$ for nearly filled states). The final factor in the above equation is the product of electronic transfer energies $t_{i,i+1}$ integrals around a closed loop of $n$ elements.

The sign of the Hall coefficient is conventional when the charge carriers are electrons ($q < 0$) moving among sites for symmetric structures in which $n$ is even. Then the Hall coefficient sign is just that of $\delta$: negative for nearly empty states ($\delta = -1$) and positive for nearly filled states ($\delta = 1$). Similarly, the sign of the entropy-of-mixing contribution to the Seebeck coefficient, proportional to $\ln[c/2(1-c)]$ or $\ln[c/(1-c)]$, is negative for a nearly empty narrow band of electrons, $c < 1/2$, and positive for a nearly full narrow electron band, $c > 2/3$.

By contrast, the sign of the Hall coefficient is often anomalous for electrons moving among sites for which $n$ is odd, as frequently occurs for non-crystalline structures. Then the sign of the Hall coefficient is just that for the product of electronic transfer energies, independent of state filling. The sign of this product only depends on the symmetry of the orbitals between which electrons transfer.

The product of $n$-transfer energies between node-less electronic states (e.g. bonding, lone-pair or ground-state impurity orbitals) is simply $(-1)^n$ because of the negative potential in orbitals' overlap regions. The sign of the Hall coefficient is then negative when $n$ is odd even while the sign of the Seebeck coefficient is positive for transport among nearly filled bonding, lone-pair or impurity-state orbitals.

Furthermore, the product of $n$-transfer energies between anti-symmetric (e.g. anti-bonding) orbitals is always positive. Thus, the sign of the Hall coefficient is positive for hopping in structures dominated by odd membered loops of antibonding orbitals. This result is opposite to the negative Hall coefficient for conventional transport involving transport among anti-bonding orbitals of a covalent semiconductor's conduction band.

Very low, $< 1$ cm$^2$/V-s, thermally activated Hall mobilities have been reported in many non-crystalline solids. Their Hall mobility activation energies are generally considerably smaller than the mobility activation energies inferred from comparison of the temperature dependences of these materials' dc conductivities and Seebeck coefficients. Hall-Effect sign anomalies are also widely reported in disordered structures that manifest hopping transport. Most strikingly, positive Hall coefficients are reported when doping introduces low-mobility hopping electrons to amorphous elemental Si and As. Negative Hall coefficients with Arrhenius Hall mobilities are also reported when dopants remove electrons from amorphous elemental Si and Ge. Unlike their crystalline counterparts, these elemental covalent semiconductors are thought to contain significant numbers of odd-membered rings. Negative Hall coefficients and positive Seebeck coefficients are widely reported for chalcogenide glasses where transport is thought to be between lone-pair orbitals on chalcogen atoms. Small-polaron-like transport between dopant-related states in heavily doped semiconductors also display negative Hall coefficients with positive Seebeck coefficients.

## Polaron absorption

Large- and small-polarons generate broad absorption bands. Features of their absorptions distinguish between them.

A large-polaron's principal absorption arises from photo-exciting its self-trapped electronic carrier into the energy band which is its genesis. The potential well binding a large-polaron's electronic carrier is approximated as that formed with surrounding atoms fixed at their carrier-induced equilibrium positions. For a Coulomb-like self-trapping potential well, the threshold for photo-ionizing its electronic carrier occurs when the photon energy $\hbar\Omega$ exceeds three times the large-polaron binding energy $3E_p$. Since $E_p >> \hbar\omega$ for a strong-coupling large-polaron, the threshold for its main absorption lies above the characteristic phonon energy $\hbar\omega$. Above this threshold the absorption increases as it mirrors the rise of the energy band's density-of-states beyond its edge. This large-polaron's absorption band subsequently declines gradually as the wave-lengths of the energy band's Bloch states fall below the self-trapped electronic carrier's spatial extent. Thus, a large-polaron primary absorption band tends to rise more sharply with increasing $\Omega$ below its peak than it declines subsequently. In addition, subsidiary narrow absorption bands can emerge from exciting self-trapped electronic carriers into their intra-well excited states.

Coherently moving charge carriers exhibit a Drude-type absorption corresponding to the real part of the frequency-dependent conductivity, $Re[\sigma(\Omega)] = \sigma(0)/[1 + (\Omega\tau)^2]$, where $\tau$ denotes the carrier's scattering time. Large polarons are extremely slow-moving, heavy-massed and weakly scattered compared with conventional electronic charge carriers. In particular, the scattering time associated with a large-polaron's scattering by acoustic phonons is so long that the large-polaron Drude absorption is confined to well below the characteristic phonon energy $\hbar\omega$. Moreover, this scattering time increases as the temperature is reduced. Therefore, the pseudo-gap between a large-polaron's low-frequency Drude absorption, $\Omega < \omega$, and the threshold of its high-frequency broad absorption band, $\Omega > \omega$, widens as the temperature is lowered. This signature behavior is unlike that of conventional electronic carriers whose Drude absorption simply falls monotonically with increasing frequency up to very high frequencies, e.g., $\Omega \approx 10^{15}$ Hz.

A small-polaron absorption is a photo-induced transition between severely localized electronic states. It elevates a small-polaron's self-trapped electron from its adiabatic ground-state into a severely localized state centered on a nearby unoccupied site. With a short-range electron-phonon interaction the absorption's peak is at twice the small-polaron binding energy. This absorption is broadened as vibrations of atoms surrounding its initial and final sites spread their electronic energies. With increasing temperature this broadening increases from that due to atoms' zero-point vibrations to that due to their thermal vibrations. The asymmetry of a small-polaron absorption band is opposite that of a large-polaron's primary high-frequency absorption band. In particular, the small-polaron absorption rises more slowly with increasing $\Omega$ below its peak than it falls above its peak.

## Interactions between polarons

Transient absorptions, recombination and possible luminescence are governed by the interactions between injected oppositely-charged carriers that form polarons. Polarons are indicated by their mobilities being well below the minimum for non-polaron carriers, $q\hbar/m_ekT$, about 300 cm²/V-s at room temperature. Since polarons are slow moving, the energy of their mutual Coulomb interaction is reduced by their material's static dielectric constant $\varepsilon_0$. Well-separated polarons also each lower their energies by their polaron binding. However, as oppositely charged polarons approach close enough to one another the pair increasingly presents itself to surrounding atoms and ions as being charge neutral. As a result, their net polaron binding energy is reduced, increasing the energy of a pair of oppositely charged polarons. In other words, there is a short-range repulsion between oppositely charged polarons. Finally, when electron and hole polarons share a common site, they simply become a self-trapped exciton. Consistent with its charge neutrality, an exciton has a much smaller energy reduction from displacement of surrounding atoms than do a separated electron and hole. These features are illustrated in Fig. 7, where, the energy of oppositely charged polarons is plotted against their separation $s$. The short-range repulsion between oppositely charged polarons is most severe in materials with especially displaceable ions, solids with exceptionally large static dielectric constants. The suppression of recombination in such materials (e.g. perovskites) fosters their functioning as especially efficient solar cells (Emin, 2018).

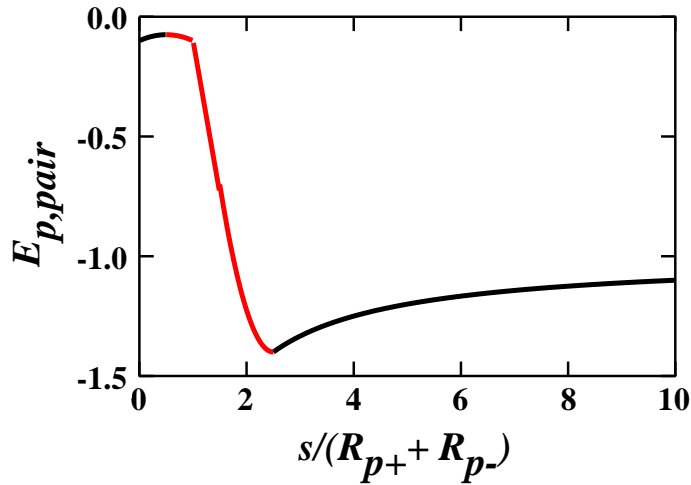

Figure 7 The energy of a pair of oppositely charged polarons $E_{p,pair}$, in units of the net energy of the pair of well-separated polarons, is plotted against their mutual separation $s$ in units of the sum of the radii of their self-trapped electronic carriers, $R_{p+}$ and $R_{p-}$. The energy reduction from relaxation of well-separated charged polarons, $s \rightarrow \infty$, is much larger than that for a self-trapped exciton, $s \rightarrow 0$. The repulsive portion of the curve is shown in red.

Repulsive interactions occur between polarons of like-charge as their self-trapped electronic carriers compete to displace intervening atoms. Self-trapping then becomes untenable at sufficiently large charge-carrier densities. This effect is the basis of a mechanism to switch a low-conductivity small-polaron semiconductor into a higher conductivity state. Specifically, current flow drives conventional electronic charge carriers from conducting leads into a semiconductor in which they form small-polarons. Small polarons diffuse much slower than conventional carriers. As a result, increasing current flow causes small polarons to progressively accumulate in the semiconductor near its interfaces with its conducting contacts. A high enough current will augment this interfacial concentration of small polarons enough to destabilize them, converting them into conventional carriers. The large interfacial concentration of small polarons thereby propagates into the semiconductor. If the propagated small-polaron concentration remains sufficiently large it too will be destabilized. The avalanche of converting small-polarons into conventional carriers will continue until the excessive small-polaron concentration is relieved (e.g., by recombination with oppositely charged carriers entering from the opposing contact). The resistance of the switched semiconductor is then primarily that of the residual region whose charge carriers remain as small polarons. The hallmark of such switching is that the resistance in the high-conductivity state is independent of the semiconductor's length since it mainly depends on the length of the residual region. As illustrated in Fig. 8, the higher conductivity state persists as long as the current is maintained above the threshold needed to initiate switching from the low-conductivity state. Furthermore, threshold switches can be converted into memory switches if their relatively large current flows induce ("burns in") structural changes into the semiconductor. Such switching has been reported for about a century in materials (e.g. chalcogenide glasses) whose very low thermally activated mobilities suggest that their charge carriers are small-polarons.

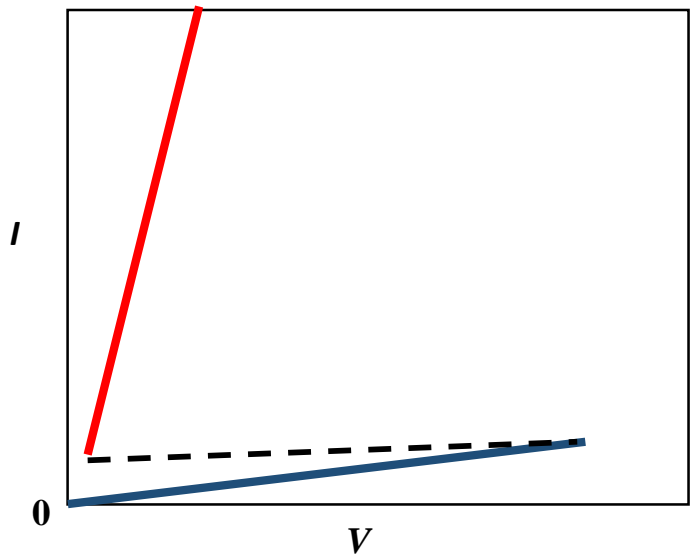

Figure 8 The current $I$ is plotted versus the voltage $V$ for a threshold switch. The low-conductivity state is shown in blue and the high-conductivity state is depicted in red. The high-conductivity state is maintained provided $I$ is kept above its threshold value.

A large singlet bipolaron has two self-trapped electronic carriers with opposing spin alignments sharing a common multi-site ground-state orbital. The Pauli principle requires that further electronic carriers added at the bipolaron's location be placed in orbitals of higher energy. This promotion energy destabilizes such multi-carrier entities with respect to separating into bipolarons and polarons. Thus, singlet bipolarons are the only multi-carrier polarons. In other words, there is a hard-core short-range repulsion between singlet bipolarons. Bipolarons also experience a mutual Coulomb repulsion. Since a bipolaron's motion is very slow, governed by associated atoms' movements, this long-range repulsion is reduced by its material's static dielectric constant, $\varepsilon_0$. This suppression is severe for large singlet bipolarons since, as noted earlier, their stability with respect to both 1) separating into two large polarons and 2) collapsing into a small bipolaron requires a material with exceptionally displaceable ions, $\varepsilon_0 > 2\varepsilon_\infty$ >> 1. The polarization of a large bipolaron's self-trapped electronic carriers in response to related atoms' vibrations reduces their energies. This softening of atoms' vibrations increases with the density of large-bipolarons. In other words, there is a phonon-mediated attraction intermediate-range between large bipolarons. The combination of large-bipolarons' mutual short-range repulsion, suppressed Coulomb repulsion and intermediate-range attraction fosters a gas of large bipolarons condensing into a large-bipolaron liquid. In addition, the liquid's density is limited in order to avoid destabilizing its large bipolarons.

**Superconductivity of a large-bipolaron liquid**

Conventional superconductivity occurs in elemental metals such as Al, Hg, Pb, La, Nb, Sn and Ta with transition temperatures below 10 K. Unconventional superconductivity occurs in doped transition-metal-oxide insulators with especially displaceable ions, as indicated by their exceptionally large values of $\varepsilon_0/\varepsilon_\infty$. The ferroelectric insulator $SrTiO_3$ is a perovskite ferroelectric whose structure can be viewed as comprising alternating planar $TiO_2$ and SrO layers. It displays superconductivity below 1 K when charge carriers are introduced by Nb substitutions or oxygen vacancies. Much higher superconducting transition temperatures are found upon doping cuprate insulators which have more complicated perovskite-based structures. For example, $La_2CuO_4$ is comprised of $CuO_2$ layers alternating with two LaO layers. Holes are introduced in the $CuO_2$ layers when divalent cations such as $Sr^{2+}$ or $Ba^{2+}$ cations are substituted for $La^{3+}$. Cuprate superconductors have superconducting transition temperatures, between about 30 K and 130 K.

The charge carriers in the normal states of conventional superconductors are quasi-free electrons. By contrast, the charge-carriers in transition-metal oxides are widely viewed as some type of polaron. In particular, polarons related to the relatively narrow bands associated with motion between transition-metal ions usually are small. Polarons associated with the relatively wide oxygen-based electronic bands are typically large. Moreover, singlet large-bipolaron formation is strongly promoted when carriers are confined to a plane embedded within a medium for which $\varepsilon_0/\varepsilon_\infty \gg 1$.

Conventional superconductivity's genesis is a phonon-mediated attraction between quasi-free electrons that induces their condensing into singlet pairs. These Cooper pairs overlap very strongly with one another to produce a quantum liquid whose ground state has an energy gap in its excitation spectrum. An unconventional model of superconductivity envisions the phonon-mediated intermediate-range attraction between large bipolarons driving their condensation into a liquid of non-overlapping charged bosons. The ground state of this liquid remains fluid, rather than condensing into a solid, if its large bipolarons do not order in a manner commensurate with its solid's underlying lattice. Large-bipolaron superconductivity is then akin to the superfluidity of a liquid of neutral mobile bosons, liquid $^4$He, but with mobile bosons of charge $2e$, a large-bipolaron liquid. As such, the liquid's long-wavelength excitations are just the large-bipolarons' plasma oscillations. These excitations' characteristic energy is $\hbar\Omega_p \equiv \hbar[4\pi(2e)^2 n_{lbp}/m_{lbp}\varepsilon_0]^{1/2}$, where $n_{lbp}$ and $m_{lbp}$ respectively denote the large-bipolaron density and effective mass. The very large values of $m_{lbp}$ and $\varepsilon_0$, which characterize large bipolarons and their formation, limit $\hbar\Omega_p$ to less than that of related phonons $\hbar\omega$. The liquid's Bose condensation and the onset of large-bipolaron superconductivity occurs when $kT$ falls below $\hbar\Omega_p$. The liquid's collective ground state is characterized by the synchronous zero-point vibrations of the atoms whose displaced-equilibrium positions generate the potential wells that self-trap the singlet pairs of its large bipolarons.

Superconductors based on doped ionic insulators display properties consistent with two-fluid large-bipolaron superconductivity. The scattering of channeled ions and the positron annihilation rate progressively fall as the homogeneous synchronous vibrating condensate grows upon lowering the temperature below the superconducting transition temperature.

Distinctive properties reported for superconductors based on doped ionic insulators are consistent with those of large bipolarons. The very high superconducting transition temperatures in cuprates may simply result from their large values of $\varepsilon_0$ ($\geq 50$) being considerably smaller than those of doped-$SrTiO_3$, a bona-fide ferroelectric, with $\varepsilon_0 \approx 20,000$ near its superconducting temperature, about 1 K.

As illustrated in Fig. 9, large-bipolarons' superconductivity requires their density to be 1) large enough to avoid their being bound to dopants, 2) small enough to preclude their destabilization, and 3) incommensurate with their ground-state's global solidification. For example, superconductivity in tetragonal $La_{2-x}Ba_xCuO_4$ only exists over a restricted doping range and disappears at an intermediate doping level that is commensurate with bipolarons' global solidification: a bipolaron centered on every fourth $CuO_2$ unit in the $x$ and $y$ directions of its planar square lattice, $2/(4\times4) = 1/8$.

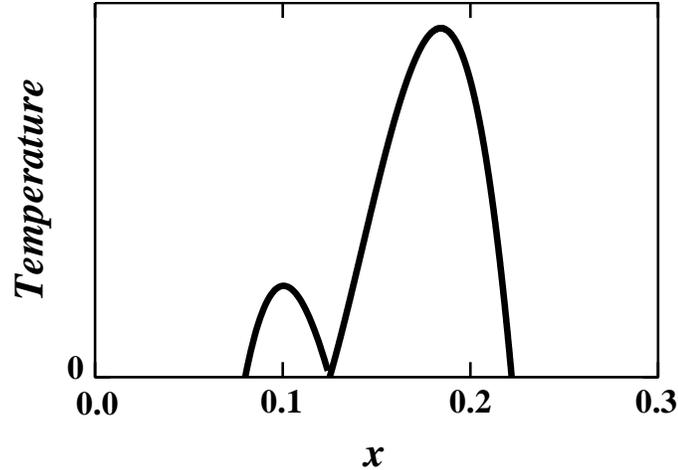

Figure 9 The superconducting transition temperature of tetragonal La$_{2-x}$Ba$_x$CuO$_4$ is plotted versus $x$, the idealized concentration of doping-induced holes. Superconductivity exists between limiting dopant concentrations. Distinctively, as observed, superconductivity also vanishes at the intermediate hole concentration $x = 1/8$. This concentration corresponds to doping-induced holes ordering as bipolarons commensurate with the basal plane's CuO$_2$ square lattice at $x = 2/(4\times4) = 1/8$.

The pseudo-gaps observed in cuprate superconductors' absorption spectra resemble the temperature-dependent gap between a large-(bi)polaron's low-frequency, $< \hbar\omega$, Drude absorption and the broad high-frequency, $> \hbar\omega$, absorptions generated by exciting its self-trapped electronic carriers. Normal-state carrier mobilities in cuprates, about 1 cm²/V-s, are much less than the minimum for coherent transport, $eh/kTm_c$, if the carrier mass $m_c$ is comparable to that of an electron. However, measured mobilities are compatible with large-(bi)polarons' huge effective masses.

As illustrated in Fig. 4, the core of a large-bipolaron in single-cuprate-layer superconductors is envisioned as two electrons being removed from the four O$^{2-}$ anions circumscribed by four spin-1/2 Cu$^{2+}$ cations residing at the corners of a CuO$_2$ planar square unit (Emin, 2017). Self-trapping occurs as the oxygen anions relax inward and the copper cations relax outward. Concomitantly, four electrons are transferred from the oxygen anions to the copper cations thereby converting them to spin-less Cu$^{1+}$ cations. The carrier-induced loss of magnetism and its spin entropy generates a negative contribution to the large-bipolaron-hole's Seebeck coefficient. It falls toward $-(k/2e)[4\ln(2)]$ with rising temperature. The superconducting state's synchronous lowest-frequency and largest-amplitude radial zero-point optic-mode vibrations of carrier-induced-displaced oxygen and copper atoms have $d$-symmetry. The adiabatic principle insures that the symmetries of these atoms' equilibrium and zero-point vibratory displacements will be reflected in their electronic states.

## Summary

An electronic charge carrier is self-trapped when it is bound within the potential well generated by displacing atoms to new equilibrium positions. Self-trapping requires that the binding energy of the self-trapped carrier exceed $h\nu$, where $\nu$ denotes the characteristic frequency of associated atoms' vibrations. The electronic carrier then circulates within the potential well that binds it before the associated atoms move appreciably. Relaxation of self-trapped electronic carriers to these atoms' vibrations lowers their frequencies. A strong-coupling polaron refers to the composite quasiparticle comprising the self-trapped electronic carrier taken together with the alterations it induces in related-atoms' equilibrium positions and vibrations.

Polaron formation is driven by a charge carrier's short-range and long-range electron-phonon interactions. These interactions result from the respective dependences of an electronic charge carrier's energy on the locations of the atoms it contacts and on displacements of distant ions. A macroscopic measure of the displaceability of these ions is the difference between the reciprocals of a material's high-frequency and static dielectric constants: $\varepsilon_\infty^{-1} - \varepsilon_0^{-1}$. For ionic and polar media $\varepsilon_0 >> \varepsilon_\infty$ but for covalent media $\varepsilon_0 \approx \varepsilon_\infty >> 1$. Carrier-induced changes in the separations between dissimilar atoms shift charge between them.

The adiabatic limit, $\nu \to 0$, presumes that electronic charge carriers always adjust to the motions of medias' relatively massive and slow moving atoms. The adiabatic potential energy is the sum of the potential energy arising from atomic displacements plus the electronic carrier as a function of atoms' positions. Adoption of a scaling argument enable the adiabatic potential to be readily studied as a function of an electronic carrier's spatial extent, its radius. A minimum of the adiabatic potential with respect to a carrier's radius indicates its being self-trapped. There are two distinct types self-trapped state. A large-polaron minimum, corresponding to a self-trapped electronic carrier extending over multiple sites, forms when long-range electron-phonon interactions predominate. A small-polaron minimum, characterized by the self-trapped electronic carrier collapsing to the most severely localized state consistent with condensed-matter's atomicity, forms when the short-range electron-phonon interaction predominates.

Self-trapped charge carriers move very slowly since their movement is contingent on atomic movements. Strong-coupling large polarons move coherently albeit with huge effective masses. As such, the minimum mobility permitted for coherently moving charge carriers ($eh/kTm_c$, where $m_c$ denotes a charge-carrier's effective mass) is much lower for large polarons than for conventional electronic charge carriers. Nonetheless, large-polaron mobilities are higher than might be expected because such heavy-massed quasi-particles tend to be difficult to scatter. Large-polaron mobilities are frequently about 1-10 cm$^2$/Vs at room temperature.

Small polarons generally move incoherently by thermally assisted hopping. Small-polaron mobilities are Arrhenius at high temperatures and become non-Arrhenius as the temperature is lowered below about $h\nu/3k$. In particular, the high-temperature adiabatic small-polaron mobility is $(ea^2\nu/kT)\exp(-E_A/kT)$, where $a$ denotes the jump distance and $E_A > h\nu$. By itself, the increase of $E_A$ with increasing jump distance favors short hops over long hops. The slowness of transfers of vibrational energy between the atoms directly related to a hop and the remainder of the medium causes hopping to occur in flurries separated by quiescent periods. An injected charge carrier's relaxation as it forms a small polaron generates a region of enhanced vibrational agitation. Its transient hopping is thereby enhanced until its full equilibration is achieved.

The Seebeck coefficient, the entropy transported with a carrier divided by its charge, displays features which distinguish polarons from conventional charge carriers. Often the change of the entropy-of-mixing generated by adding a charge carrier provides the dominant contribution to the Seebeck coefficient. Then the Seebeck coefficient just depends on the carrier concentration. By contrast, the dc conductivity also depends on charge-carriers' mobility. Taken in tandem, measurements of the Seebeck coefficient and the dc conductivity permit determination of whether the carrier mobility manifest the Arrhenius high-temperature behavior characteristic of small-polaron hopping. The Seebeck coefficients of large (bi)polarons are also distinctive in that they have contributions from carrier-induced reductions of related atoms' vibration frequencies.

Distinctively, the Hall mobility, the mobility determined from a Hall-Effect measurement, for small-polaron hopping generally rises more slowly with increasing temperature than does the mobility which enters into the dc conductivity. Moreover, the sign of the small-polaron Hall Effect is often opposite that were the charge carriers conventional. These anomalous situations occur in many amorphous semiconductors, for semiconductors' impurity conduction, and for some crystalline structures.

Since large polarons are massive and relatively difficult to scatter, the declines of the real part of their conductivities with increasing applied frequency primarily occur well below those of atomic vibrations, $10^{13}$ Hz, rather than at the relative high frequencies, $10^{15}$ Hz, characterizing conventional electronic charge carriers. In addition, the large-polaron frequency-dependent conductivity associated with exciting its self-trapped electron carrier occurs above atomic vibration frequencies. Thus, unlike a conventional electronic carrier, a large-polaron's absorption spectrum has a low-energy gap which widens with decreasing temperature. Small polarons' absorption spectra are also qualitatively distinct from those of conventional electronic charge carriers. The photon-assisted inter-site transfer of a small-polaron generates a broad absorption band which peaks at an energy of about four times the hopping activation energy.

Distinctive features of (bi)polarons' mutual interactions support some device applications. In particular, as oppositely charged polarons approach one another their net interaction with distant ions progressively diminishes. The net polaron-formation energy is thereby reduced. In other words, there is a short-range repulsion between oppositely charged polarons (Emin, 2018). By suppressing recombination of oppositely charged large polarons in ionic solids this effect fosters their solar cell efficiency. This effect may account for the high efficiencies of perovskite solar cells whose 1) very large values of $\varepsilon_0/\varepsilon_\infty$, 2) moderate carrier mobilities (i.e., about 1 cm$^2$/V-s at 300 K), and 3) distinctive absorption spectra suggest large-polaron formation. This effect would also increase the feasibility of high-efficiency, long-lived, high-power, beta-voltaic devices (Emin, 2019). These devices are stacks of p-n junctions of doped icosahedral boride semiconductors in which a reliable, high intensity beta-particle flux from a radio-isotope (e.g., $^{90}$Sr) replaces a solar-cell's intermittent moderate-intensity solar flux.

The density of stable polarons is limited by their competition to displace surrounding atoms. Thus, introducing ever higher concentrations of polarons into a semiconductor ultimately destabilizes them with respect to their forming conventional high-mobility electronic charge carriers. Thus, driving a large enough current through a semiconductor whose equilibrated carriers are small polarons cause it to switch into a low resistance state. This low-resistance state is maintained by keeping the current above the threshold value at which switching in initiated.

Charge carriers can pair as singlet large-bipolarons in materials with exceptionally displaceable ions as indicated by $\varepsilon_0/\varepsilon_\infty \gg 1$. The phonon-mediated attraction between large bipolarons, generated by their self-trapped electrons' collective response to atoms' vibrations, fosters their condensing into a liquid. Superconductivity is associated with the Bose condensation of this large-bipolaron liquid. Distinctive features distinguish this superconductivity from metals' conventional superconductivity. In particular, 1) large bipolarons and their liquid only exist for moderate carrier concentrations, 2) the normal-state large-bipolaron mobility is much less than that permitted for conventional electronic charge carriers, and 3) large-bipolarons' normal-state absorption spectra have low-energy gaps that widen with decreasing temperature.

Electronic charge carriers that self-trap, thereby forming strong-coupling (bi)polarons, display transport and optical properties which distinguish them from conventional carriers. Large polarons display exceptionally low mobilities that decrease with increasing temperature. A large-polaron's absorption spectra consist of a broad high-frequency band that is separated by a pseudo energy gap from an extremely narrow low-frequency Drude-type absorption. Small polarons manifest thermally-assisted mobilities which are usually very much lower than even large-polaron mobilities. Distinctively, small-polaron mobilities become Arrhenius at high temperatures. A small-polaron's Hall-Effect mobility increases with rising temperature more gently than that of the mobility which enters into the dc conductivity. In many materials a magnetic field deflects a small polaron in the opposite direction as it would deflect a free carrier of the same charge. A broad small-polaron absorption band is associated with photon-assisted hopping. Oppositely charged polarons experience mutual short-range repulsion which impede their recombination. Polarons of like charge can combine to form singlet bipolarons in materials with exceptionally displaceable ions. Large bipolarons may even condense into a liquid which undergoes a Bose condensation to yield unconventional superconductivity. Reports of most of these extraordinary effects already exist. Observations of the unusual properties which identify strong-coupling polarons may become more common as increasingly complex semiconductors are investigated.

**References**

Emin, D., 2006. Unusual properties of icosahedral boron-rich solids. Journal of Solid State Chemistry. 179, 2791-2798.

Emin, D., 2013. Theory of Meyer-Neldel compensation for adiabatic charge transfer. Monatshefte für Chemie – Chemical Monthly. 144, 3-10.

Emin, D., 2016. Determining a hopping polaron's bandwidth from its Seebeck coefficient: measuring the disorder energy of a non-crystalline semiconductor. Journal of Applied Physics. 119, 045101.

Emin, D., 2017. Dynamic d-symmetry Bose condensate of a planar-large-bipolaron-liquid in cuprate superconductors. Philosophical Magazine. 97, 2931-2945.

Emin, D., 2018. Barrier to recombination of oppositely charged large polarons. Journal of Applied Physics. 123, 055105.

Emin, D., 2019. Proposed high-capacity efficient energy cells from $MgAlB_{14}$-type icosahedral-boron semiconductors. AIP Advances. 9, 055226.

**Further Reading**

Emin, D, 2013. Polarons. Cambridge University Press, Cambridge.